\documentclass[aps,pra,superscriptaddress,amsmath,amsfonts,amssymb,twocolumn,showpacs]{revtex4}
\usepackage{wasysym}
\renewcommand{\vec}[1]{\mathbf{#1}} \usepackage{graphicx}
\usepackage{color}

\newcommand{\figref}[1]{Fig.~\ref{fig:#1}}
\newcommand{\Figref}[1]{Figure~\ref{fig:#1}}

\renewcommand{\eqref}[1]{Eq.~\ref{eq:#1}}

\newcommand{\citeasnoun}[1]{Ref.~\onlinecite{#1}}

\definecolor{Orange}{cmyk}{0,0.61,0.87,0}
\definecolor{BrickRed}{cmyk}{0,0.89,0.94,0.28}
\definecolor{DarkGreen}{rgb}{0,0.7,0.1}
\definecolor{Vio}{rgb}{0.509,0.248,0.987}
\def\a{s}
 \def\b{s}

\newcommand{\addn}[1]{\if\a\b{{\color{red} #1}}\else{#1}\fi}
\newcommand{\add}[1]{\if\a\b{{\color{magenta} #1}}\else{#1}\fi}
\newcommand{\rjadd}[1]{\if\a\b{{\color{DarkGreen}RJ: #1}}\else{#1}\fi}
\newcommand{\del}[1]{{\if\a\b{{\color{black}[[#1]]}}\else{}\fi}}
\newcommand{\rjdel}[1]{{\if\a\b{{\color{BrickRed}[[RJ: #1]]}}\else{}\fi}}

\newcommand{\commf}[1]{\if\a\b{{\color{Vio}\{\small  \sc TE: #1\}}}\else{}\fi}
\newcommand{\comm}[1]{}
\newcommand{\commn}[1]{}
\newcommand{\rjcomm}[1]{}
\newcommand{\commte}[1]{}

\begin{document}

\title{Nonmonotonic effects of parallel sidewalls on Casimir forces between cylinders}

 \author{Sahand~Jamal~Rahi}
  \affiliation{Department of Physics,
  Massachusetts Institute of Technology, Cambridge, MA 02139}
  \author{Alejandro W. Rodriguez}
  \affiliation{Department of Physics,
  Massachusetts Institute of Technology, Cambridge, MA 02139}
  \author{Thorsten Emig}
  \affiliation{Institut f\"ur Theoretische Physik, Universit\"at zu K\"oln,
Z\"ulpicher Strasse 77, 50937 K\"oln, Germany}
  \affiliation{CNRS, LPTMS, UMR8626, B\^at.~100, Universit\'e Paris-Sud, 91405 Orsay, France}
  \author{Robert L.~Jaffe}
  \affiliation{Center for Theoretical Physics and Laboratory for Nuclear Science, Massachusetts Institute of Technology, Cambridge, MA 02139}
  \author{Steven G. Johnson}
  \affiliation{Department of Mathematics,
  Massachusetts Institute of Technology, Cambridge, MA 02139}
   \author{Mehran Kardar}
  \affiliation{Department of Physics,
  Massachusetts Institute of Technology, Cambridge, MA 02139}

\begin{abstract}

We analyze the Casimir force between two parallel infinite metal
cylinders, with nearby metal plates (sidewalls), using
complementary methods for mutual confirmation.  The attractive force
between cylinders is shown to have a \emph{nonmonotonic} dependence on
the separation to the plates. This intrinsically multi-body
phenomenon, which occurs with either one or two sidewalls
(generalizing an earlier result for squares between two sidewalls),
does not follow from any simple two-body force description. 
 We can, however, explain the nonmonotonicity by considering the screening
(enhancement) of the interactions by the fluctuating charges
(currents) on the two cylinders, and their images on the nearby
plate(s). Furthermore, we show that this effect also implies a
nonmonotonic dependence of the cylinder-plate force on the
cylinder-cylinder separation.

\end{abstract}


\maketitle

\begin{figure}[t]
\includegraphics[width=0.50\columnwidth]{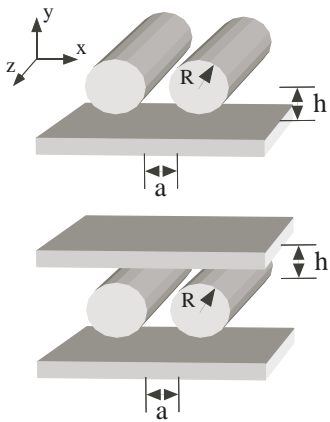}
\caption{Schematic of the cylinder-plate geometry consisting of two
infinite metallic cylinders of radius $R$, separated by a distance $a$
from each other, and at a distance $h$ from one (top) or two (bottom)
metallic sidewall(s).
\commte{Labels a, h seem to be shifted, i.e., do not match with drawing.}
}
\label{fig:geom}
\end{figure}

Casimir forces arise from quantum vacuum fluctuations, and have been
the subject of considerable theoretical and experimental
interest~\cite{casimir,Lifshitz80,Onofrio06,Capasso07:review}. 
We consider here the force between metallic
cylinders with one or two parallel metal sidewalls (\figref{geom})
using two independent exact computational methods, 
and find an unusual nonmonotonic dependence of the force on the sidewall
separation.  These nonmonotonic effects cannot be predicted by
commonly used two-body Casimir-force estimates, such as the
proximity-force approximation (PFA)~\cite{Bordag06,Parsegian06} that
is based on the parallel-plate limit, or by addition of Casimir-Polder
`atomic'
interactions (CPI)~\cite{Casimir48:polder,Sedmik06,Parsegian06}.
\comm{Add sentence on works that in the past have demonstrated the
failure of PFA in various limits [cite gies. et. al,
~\citeasnoun{gies06:edge}]}

In previous work, we demonstrated a similar nonmonotonic force between
two metal squares in proximity to two parallel metal sidewalls, for
either perfect or realistic metals~\cite{Rodriguez07:PRL}.  This work,
with perfect-metal cylinders~\cite{emig06}, demonstrates that the
effect is not limited to squares (i.e., it does not arise from sharp
corners or parallel flat surfaces), nor does it require two sidewalls.
The nonmonotonicity stems from a competition between forces from
transverse electric (TE) and transverse magnetic (TM) field
polarizations: In the latter case, the interaction between fluctuating
charges on the cylinders is screened by opposing image charges, in the
former case it is enhanced by analogous fluctuating image
currents. Furthermore, we show that a related nonmonotonic variation
arises for the force between the cylinders and a sidewall as a
function of separation between the cylinders, a geometry potentially
amenable to experiment.

Casimir forces are not two-body interactions: quantum fluctuations in
one object induce fluctuations throughout the system which in turn act
back on the first object.  However, both the PFA and CPI view Casimir
forces as the result of attractive two-body (``pairwise'')
interactions. 
They are reasonable approximations only in certain
limits (e.g., low curvature for PFA), and can fail qualitatively as
well as quantitatively otherwise.
Pairwise
estimates fail to account for two important aspects of the Casimir
forces in the geometry we consider~\cite{Zaheer07}.  First, 
a monotonic pairwise attractive force clearly cannot give
rise to the nonmonotonic effect of the sidewalls.  Second, the
application of either method here would include two contributions to
the force on each cylinder: attraction to the other cylinder
and attraction to the sidewall(s).  If the latter contribution is
restricted to the portion of the sidewall(s) where the other cylinder
does not interpose (``line of sight'' interactions), the
cylinder will experience greater attraction to the 
opposite side, thereby \emph{reducing} the
net attractive force between the cylinders~\cite{Zaheer07}.  In
contrast, exact calculations predict a nonmonotonic force that is
\emph{larger} in the limit of close sidewalls than for no sidewalls.
These important failures illustrate the need for caution when applying
uncontrolled approximations to new geometries even at the qualitative
level.  (On the other hand, a ray-optics
approximation, which incorporates non-additive many-body effects,
at least qualitatively predicts these features for the case of
two-square/sidewall~\cite{Zaheer07}.)  \rjcomm{Following
Thorsten's recommendation, I suggest putting a slightly elaborated
version of the parenthesis (``line of sight'') that follows into a
footnote}

\Figref{geom} is a schematic of the three-dimensional geometries that
we consider: two infinite, parallel, perfect-metal cylinders of radius
$R$ separated by a distance $a$ (center-to-center separation $2R+a$)
and oriented along the $z$ axis, with one (top panel) or two (bottom
panel) infinite metal sidewall(s) parallel to the cylinders and
separated from both by equal distance $h$ ($h+R$ to the cylinder axes). 
For perfectly conducting objects with $z$-translational symmetry, the electric
($\vec{E}$) and magnetic ($\vec{H}$) fields can be decomposed into TE
and TM polarizations, described by scalar fields $\psi$ satisfying
Neumann (TE, $\psi=H_z$) and Dirichlet (TM, $\psi=E_z$) boundary
conditions at the metallic surfaces~\cite{Collin91}.

To analyze these geometries, we employ two complementary and exact
computational methods, based on path integrals (PI) and the mean
stress tensor (ST).  The methods are \emph{exact} as they involve
no uncontrolled approximations and can yield arbitrary accuracy given
sufficient computational resources.  They are \emph{complementary} in
 that they have different strengths and weaknesses.  The PI
method is most informative at large separations where it leads to
analytical asymptotic expressions. The ST method, while relatively
inefficient for large separations or for the specific geometries where
PI is exponentially accurate, is formulated in a generic fashion that
allows it to handle arbitrary complex shapes and materials without
modification. As both of methods are described in detail elsewhere
~\cite{Rodriguez07:PRA, emig05},  we only summarize them briefly
here.  The present geometry provides an arena where both
methods can be applied and compared.

In the PI approach, the Casimir force is calculated via the
constrained partition function. The Dirichlet (Neumann) constraints on
the TM (TE) fields are imposed by auxiliary fields on the metallic
surfaces~\cite{emig03_1} which can be interpreted as fluctuating
charges (currents).  The interaction between these charges is related
to the free-space Green's functions---the addition of metallic
sidewall(s) merely requires using image charges (currents) to enforce
the appropriate boundary conditions. The calculations are further
simplified by using Euclidean path integrals and the corresponding
imaginary-frequency $\omega = iw$ Green's function.  In the 
case of infinite cylinders, these surface fields can be represented in
terms of a spectral basis: their Fourier series, leading to Bessel
functions in the Green's function~\cite{emig06, Bordag06,
Dalvit06}. An important property of such a spectral basis is that its
errors go to zero exponentially with the number of degrees of freedom.

\begin{figure}[t]
\includegraphics[width=1.0\columnwidth]{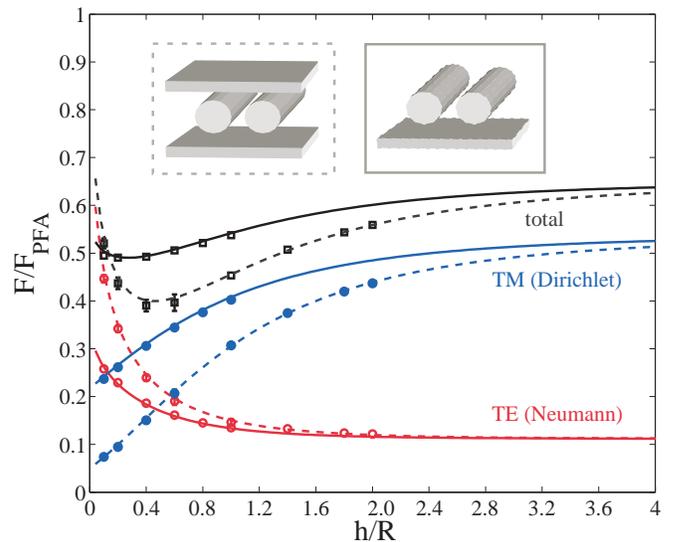}
\caption{Casimir force per unit length between two cylinders (black)
  vs. the ratio of sidewall separation to cylinder radius $h/R$, at
  fixed $a/R=2$, normalized by the total PFA force per unit length
  between two isolated cylinders [$F_{\text{PFA}}= \frac{5}{2}(\hbar
  c\pi^3/ 1920)\sqrt{R/a^7}$~\cite{Rahi07}]. The solid lines refer to
  the case with one sidewall, while dashed lines depict the results
  for two sidewalls, as indicated by the inset.  Also shown are the
  individual TE (red) and TM (blue) forces.}
\label{fig:two-plates}
\end{figure}

We also use a method based on integration of the mean ST, evaluated in
terms of the imaginary-frequency Green's function via the
fluctuation-dissipation theorem~\cite{Rodriguez07:PRA}.  The Green's
function can be evaluated by a variety of
techniques~\cite{Rodriguez07:PRA}, but here we use a simple
finite-difference frequency-domain method (FDFD)~\cite{Christ87} that
has the advantage of being very general and simple to implement at the
expense of efficiency---it is much less efficient for this specific
geometry than the PI method.  The results from both methods are
shown in \figref{two-plates}, with the PI method indicated by
solid/dashed lines for two/one sidewalls, and the ST method indicated
by data points.  Both results agree to within the numerical accuracy,
as expected, although we have fewer data points (and larger error
bars) from the ST method because it is less efficient for this
geometry.  
We focus on the
interpretation of the results rather than on the computational
techniques. 
\rjcomm{Since there is considerable emphasis on the comparison of the
two methods, it might be useful to give the reader some indication of
the computational time required to obtain the results shown in
Fig. 2.}

To begin with, we compute the force between the two cylinders (with
$a/R= 2$) as a function of the sidewall separation $h/a$, for fixed
$a$.  The results, for both one sidewall (solid lines) and two
sidewalls (dashed lines) are shown in \figref{two-plates} for the
total force (black), and also the forces for the individual
polarizations TE (red) and TM (blue).  
The forces are all
normalized  to the PFA result between two
cylinders~\cite{Rahi07}, which is independent of $h$ and 
does not affect the shape of the curves
in \figref{two-plates}. 
Two interesting observations can be made from this figure.  First, the total force (for both one and two sidewalls)
is a nonmonotonic function of $h/R$: at first decreasing and then
increasing towards the asymptotic limit  between two
isolated cylinders.
 The extremum for the one-sidewall case occurs at $h/R \approx 0.27$,
and for the two-sidewall case is at $h/R \approx 0.46$ (similar to the
$h/R \approx 0.5$ for squares between two
sidewalls~\cite{Rodriguez07:PRL}).  Second, the total force for the
two-sidewall case in the $h=0$ limit is larger than for
$h\rightarrow\infty$. As might be expected, the $h$-dependence for one
sidewall is weaker than for two sidewalls, and the effects of the two
sidewalls are not additive: not only is the difference from
$h\rightarrow\infty$ force not doubled for two sidewalls compared to
one, but the two curves actually intersect at one point.  

Since nonmonotonic sidewall effects appear to occur for a variety of
shapes (both for square~\cite{Rodriguez07:PRL} and circular cross
sections), it is natural to seek a simple generic argument to explain
this phenomenon.  As we see in \figref{two-plates}, and also in our
earlier work~\cite{Rodriguez07:PRL}, the nonmonotonicity arises from a
competition between the TE and TM force contributions: the
TE force is quickly decreasing with $h$ while the TM force is slowly
increasing. Therefore, an
underlying question is why the TE force increases as the sidewall
comes closer, while the TM force decreases.

\begin{figure}[t]
\includegraphics[width=1.0\columnwidth]{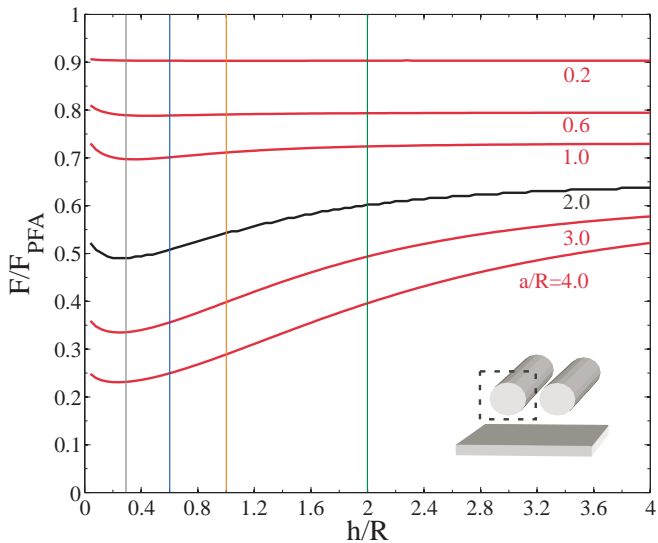}
\caption{Casimir force per unit length between two cylinders of fixed
  radius $R$ vs. the ratio of sidewall separation to cylinder radius
  $h/R$ (for one plate), normalized by the total PFA force per unit
  length between two isolated cylinders. The force is plotted for
  different cylinder separations of $a/R$ = $0.2$, $0.6$, $1.0$, $2.0$,
  $3.0$, and $4.0$.}
\label{fig:oneplate}
\end{figure}

\begin{figure}[t]
\includegraphics[width=1.0\columnwidth]{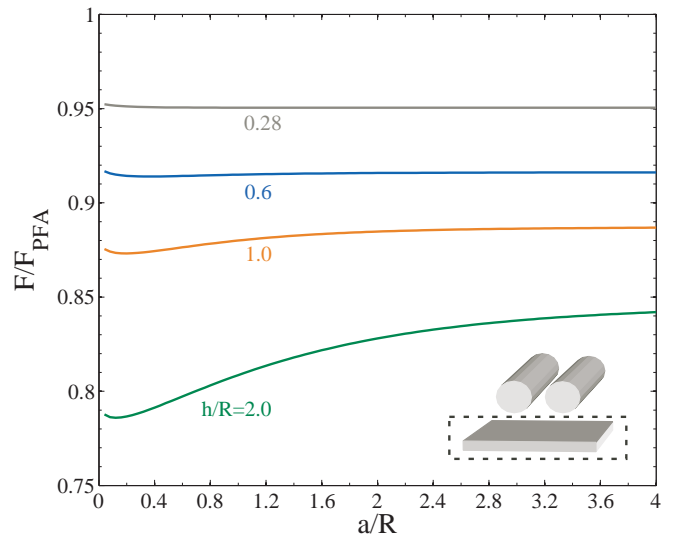}
\caption{Casimir force per unit length between a plate and two
  cylinders of fixed radius $R$ vs. the ratio of cylinder separation
  to cylinder radius $a/R$, normalized by the total PFA force per unit
  length between a cylinder and a plate $F_{\text{PFA}}=
  \frac{5}{2}(\hbar c\pi^3/ 960)\sqrt{R/2a^7}$~\cite{Rahi07}. The
  force is plotted for different plate separations of $h/R$ = $0.28$,
  $0.6$, $1.0$, and $2.0$.  Note that the normalization is
  different from the cylinder-cylinder PFA in the previous figures.
  \commte{Normalization should be with respect to PFA for
  plate-cylinder. Is it this what has been done here? Then, the
  different meaning of $F_{PFA}$ in Figs. 3 and 4 should be mentioned
  in the text also.}  }
\label{fig:cylplate}
\end{figure}

An intuitive perspective of the effects of the metallic sidewall(s) on
the TE/TM forces is obtained from the ``method of images,''
whereby the boundary conditions at the plate(s) are enforced by
appropriate image sources.  For the Dirichlet boundary conditions (TM
polarization) the image charges have \emph{opposite} signs, and
the potential due to a charge (more precisely, the Green's function at
any imaginary frequency, which determines the Casimir force) is
obtained by subtracting the contribution from the opposing image. Any
configuration of fluctuating TM charges on one cylinder is thus
screened by images, more so as $h$ is decreased, \emph{reducing} the
force on the fluctuating charges of the second cylinder~\footnote{A
key fact is that the Green's functions in Casimir forces are naturally
evaluated at imaginary frequencies~\cite{Rodriguez07:PRA}, which means
that they are decaying and not oscillating.  If they 
were oscillating, one could not as easily infer whether opposite-sign
image currents add or subtract.}. 
Since the reduction in
force is present for every configuration, it is there also for the
average over all configurations, accounting for the variations of the
TM curves in \figref{two-plates}.  By contrast, the Neumann boundary
conditions (TE polarization) require image sources (current loops) of
the \emph{same} sign.  The total force between fluctuating currents on
the cylinders is now larger and increases as the plate separation $h$
is reduced.  (An analogous additive effect occurs for the classical
force between current loops near a conducting plane.)  

Note, however,
that while for each fluctuating source configuration, the effect of
images is additive, this is no longer the case for the average over
all configurations. More precisely, the effect of an
image source on the Green's function is not additive because of
feedback effects: the image currents change the surface current
distribution, which changes the image, and so forth.  For example,
the net effect of the plate on the Casimir TE force \emph{is not} to
double the force as $h\to0$.  The increase is in fact larger than two
due to the correlated fluctuations.

While the method of images explains the competition between
TE and TM modes that underlies the nonmonotonic effects, 
further considerations are required to ensure that their sum
is nonmonotonic. For example, if the
TE and TM variations 
with $h$ were equal and opposite, they
would cancel with no net dependence on $h$.  That
this is not the case can be checked by examining the limit $a\gg h\gg
R$.  In this limit the forces are dominated by the lowest spectral
(Fourier) mode, $s$-wave for TM and $p$-wave for TE~\cite{emig06}. The
former is stronger and leads to an asymptotic form (for one plate):
\begin{equation}\label{asymptote}
\frac{F}{L} = - \frac{4 \hbar c}{\pi}\frac{  h^4}{a^7 \ln^2(R/h)},
\end{equation}
\commte{This result is for one plate, right? Should be mentioned.
Also, it should be said how many images (one?) are taken into account
which is important for argument in favor of $h^4$ scaling, see below.}
confirming the reduced net force as the cylinders approach the plate.
While the logarithmic dependence on $R$ could have been
anticipated~\cite{emig06}, the $h^4$ scaling is a remarkable
consequence of the multi-body effect.  Each cylinder and its mirror
image can be considered as a dipole of size $\sim h$. The interaction
of the two dipoles should scale as the interaction between two
cylinders of size $\sim h$ with Neumann boundary conditions.
For $a\gg h$ the force for
the latter problem scales as $\sim h^4/a^7$, explaining the above
result~\cite{emig06}, up to the logarithm. 
To analytically establish the nonmonotonic
character, we also need to show that the TE force is dominant in the
opposite limit of $h\ll R$.  So far, we only have numerical arguments
in favor of this~\cite{Rahi07}.

In \figref{oneplate}, we show the total force vs. $h/R$ for a variety
of different values of the cylinder separation $a$ in the presence of
a single sidewall.  The value $a/R = 2$ corresponds to our previous
results in \figref{two-plates}.  
Note that if $a$ is too large or too small, the
degree of non-monotonicity (defined as the difference between the
minimum force and the $h=0$ force) decreases.  (For small $a$, the
force is accurately described by PFA,
while for large $a$ the TM mode dominates as
indicated in Eq.~(\ref{asymptote}).)  The separation $a/R = 2$ from
\figref{two-plates} seems to achieve the largest value of
non-monotonicity.

When the force between the cylinders is not monotonic
in $h$, it also follows that the force between the cylinders and the
sidewalls is not monotonic in $a$.  A nonmonotonic force $F_x$ between
the cylinders means that there is a value of $h$ where $dF_x/dh$ =
0.  Since the force is the derivative of the energy, $F_x =
-\partial\mathcal{E}/\partial a$, at this point
$\partial^2\mathcal{E}/\partial a \partial h = 0$.  These two
derivatives, of course, can be interchanged to obtain
$\partial(\partial\mathcal{E}/\partial h)/\partial a = 0$.  But this
means that $dF_y/da = 0$ at the same point, where $F_y =
-\partial\mathcal{E}/\partial h$ is the force between the cylinders
and the sidewall.  This cylinders-sidewall force is plotted in
\figref{cylplate} as a function of $a/R$ for various values of $h/R$.
(It is not surprising that the effect of a small cylinder on the force
between two bodies is smaller than the effect of an infinite plate.
This is also reflected in the fact that the cylinder-cylinder force is
generally less than the cylinder-plate force for the same cylinder
diameter and surface separation~\cite{Rahi07}.)  \comm{I am not sure
if the statement in parenthesis is a proper comparison.  After all we
have scaled all forces by PFA, and the results for this case are
actually closer to PFA.} \commn{The only point here is that, for a
given cylinder diameter and surface separation, Casimir-type forces on
the cylinder (in absolute units, not scaled by PFA) generally seem to
be larger for a plate than for another cylinder, which doesn't seem
surprising given the disparity in surface areas etcetera.  The fact
that our forces here are closer to PFA does not seem relevant, since
here we are talking about the magnitude of the three-body nonmonotic
effect.}  \commte{It was not clear to me if the PFA for the
cylinder-plate is used here -- which is what should be the
case. Perhaps this can be said more clearly.}

The advantage of the cylinder-plate force compared to the
cylinder-cylinder force is that it seems operationally closer to the sphere-plate
geometries that have been realized experimentally.  In order to
measure the cylinder-cylinder force, one would need to suspend two
long cylinders in vacuum nearly parallel to one another. To measure
the cylinder-plate force, the cylinders need not be separated by
vacuum---we expect that a similar phenomenon will arise if the
cylinders are separated by a thin dielectric spacer layer.
Unfortunately, the nonmonotonic effect in \figref{cylplate} is rather
small (roughly 0.2\%), but it may be possible to increase this by
further optimization of the geometry.  In future calculations, we also
hope to determine whether the same phenomenon occurs for two spheres
next to a metal plate.

In previous research,  unusual Casimir force
phenomena were sought by considering parallel plates with exotic materials: for
example, repulsive forces were predicted using magnetic
conductors~\cite{Kenneth02}, combinations of different
dielectrics~\cite{Imry05}, fluids between the plates~\cite{Munday07},
and even negative-index media with gain~\cite{Leonhardt07}.  A
different approach is to use ordinary materials with more complicated
geometries: as illustrated in this and previous~\cite{Rodriguez07:PRL}
work, surprising nonmonotonic (attractive) effects can arise by
considering as few as three objects.

This work was supported in part by NSF grant DMR-04-26677 (SJR and
MK), by the US Dept.~of Energy (DOE) under cooperative research
agreement DF-FC02-94ER40818 (RLJ), by DOE
grant DE--FG02-97ER25308 (AR), and by 
DFG grant EM70/3 (TE).

\bibliographystyle{apsrev}
\bibliography{photon}

\end{document}